\documentclass[12pt,letterpaper]{article}
\usepackage{jheppub}
\usepackage{empheq}
\RequirePackage{pgfopts}[2011/06/02]
\usepackage{amssymb}
\usepackage{color}
\usepackage{slashed}
\usepackage{amsmath}
\usepackage{amsfonts}
\usepackage{amssymb}
\usepackage{graphicx,subcaption}
\usepackage{caption}
\usepackage{subcaption}
\usepackage{float}
\usepackage{tikz}
\usetikzlibrary[arrows]
\usepackage{amsmath}
\usepackage{epsfig}
\usepackage{appendix}
\usepackage{listings}
\usepackage{xcolor}
\usepackage{empheq}
\newcommand{\vev}[1]{{\left< {#1} \right>}}
\newcommand{\be}{\begin{equation}}
\newcommand{\ee}{\end{equation}}

\def\bZ {\mathbb{Z}}

\def\bC {\mathbb{C}}

\newcommand{\tlambda}{\tilde \lambda}
\newcommand{\traza}[1]{{\mathrm{Tr}\,a^{#1}}}

\title{The planar limit of ${\cal N}=2$ superconformal quiver theories}

\author{Bartomeu Fiol,}
\author{Jairo Mart\'inez-Montoya}
\author{and Alan Rios Fukelman}

\affiliation{Departament de F{\'\i}sica Qu\`antica i Astrof\'isica i \\Institut de Ci{\`e}ncies del Cosmos, 
Universitat de Barcelona,
Mart{\'\i}\ i Franqu{\`e}s 1, 08028 Barcelona, Catalonia, Spain}

\emailAdd{bfiol@ub.edu}
\emailAdd{jmartinez@icc.ub.edu}
\emailAdd{ariosfukelman@icc.ub.edu}

\abstract{We compute the planar limit of both the free energy and the expectation value of the $1/2$ BPS Wilson loop for four dimensional ${\cal N}=2$ superconformal quiver theories, with a product of SU($N$)s as gauge group and  bi-fundamental matter. Supersymmetric localization reduces the problem to a multi-matrix model, that we rewrite in the zero-instanton sector as an effective action involving an infinite number of double-trace terms, determined by the relevant extended Cartan matrix. We find that the results, as in the case of $\mathcal{N}=2$ SCFTs with a simple gauge group, can be written as sums over tree graphs.  For the $\widehat{A_1}$ case, we find that the contribution of each tree can be interpreted as the partition function of a generalized Ising model defined on the tree; we conjecture that the partition functions of these models defined on trees satisfy the Lee-Yang property,  {\em i.e.} all their zeros lie on the unit circle.}

\begin{document}
\maketitle
\section{Introduction}

The emergence of quantum gravity  from a gauge theory is one of the most fascinating issues that can be addressed with the AdS/CFT correspondence. Since the work of \cite{Henningson:1998gx} it has been clear that not every conformal field theory (CFT) in the large $N$ limit can be dual to a gravitational theory described by a two derivative Einstein-Hilbert action. For instance, for four dimensional CFTs a necessary condition is that the two central charges coincide in the large $N$ limit, $a=c$ \cite{Henningson:1998gx}. For instance, this property is satisfied by ${\cal N}=4$ super Yang-Mills, but it is not satisfied by ${\cal N}=2$ SU($N$) with $n_F=2N$ hypermultiplets in the fundamental representation, thus ruling out that the large $N$ limit of this CFT has a holographic dual well described by gravity.

Since the early days of the holographic correspondence, it has been important to find further examples of CFTs with holographic duals, beyond the original example of ${\cal N}=4$ SYM. Four dimensional quiver gauge theories with ${\cal N}=2$ superconformal symmetry satisfy an ADE classification \cite{Katz:1997eq}, and for certain values of the marginal couplings, they are orbifolds of ${\cal N}=4$ SYM and have a gravity dual \cite{Kachru:1998ys, Lawrence:1998ja}. These quiver gauge CFTs constitute thus an interesting laboratory, as variation of their marginal couplings allows to connect CFTs with and without gravity duals in the large $N$ limit \cite{Gadde:2009dj, Gadde:2010zi, Rey:2010ry, Pomoni:2011jj, Mitev:2014yba, Mitev:2015oty,Gadde:2012rv, Zarembo:2020tpf}.

In this work we will consider ${\cal N}=2$ SCFTs with gauge group a product of SU($N$)s, paying special attention to the simplest case, the $\widehat{A_1}$ theory, with gauge group SU($N$)$\times $SU($N$). This theory has two marginal couplings $(g_1,g_2)$ and varying them one can reach an orbifold of ${\cal N}=4$ SYM and ${\cal N}=2$ SU($N$) SQCD. Our main technical tool will be supersymmetric localization \cite{Pestun:2007rz}. Thanks to this tool,  the planar free energy and expectation value of the 1/2 BPS circular Wilson loop are known to all orders in the 't Hooft coupling for the limiting theories mentioned above  ${\cal N}=4$ SYM  and ${\cal N}=2$ SU($N$) SQCD \cite{Erickson:2000af, Drukker:2000rr, Passerini:2011fe,Russo:2012ay, Fiol:2018yuc,Fiol:2020bhf}.

Four dimensional ${\cal N}=2$ quiver CFTs have already been studied using localization \cite{Rey:2010ry,Mitev:2014yba, Mitev:2015oty,Pini:2017ouj,Zarembo:2020tpf}. The novelty of this work is that we evaluate various quantities of these theories in the planar limit, to all orders in the 't Hooft couplings $\lambda_i$. We do so by applying the same strategy developed for CFTs with simple gauge groups in \cite{Fiol:2020bhf}.  For these quiver CFTS, supersymmetric localization \cite{Pestun:2007rz}  reduces the evaluation of various quantities to matrix integrals.  Compared to the case of ${\cal N}=2$ SCFTs with a simple gauge group, the main novelty is that the resulting matrix models are multi-matrix models. In the simplest case, the model to solve is a two-matrix model. As in our recent work  \cite{Fiol:2020bhf}, we rewrite the 1-loop factor as an effective action involving an infinite number of double-trace terms, in the fundamental representation of the respective gauge groups. We then show that this double-trace form of the potential implies that the perturbative series considered admit a combinatorial formulation, as sums over tree graphs.

While we will present results valid for all ${\cal N}=2$ quiver CFTs, we will pay special attention to the simplest theory, $\widehat{A_1}$. This theory has a $\bZ_2$ symmetry exchanging the two nodes of the quiver. Since the ranks of the gauge groups are equal, this $\bZ_2$ symmetry amounts to exchanging $g_1\leftrightarrow g_2$. We will be particularly interested in observables that transform nicely under this symmetry: the free energy and particular linear combinations of the usual 1/2 BPS circular Wilson loop defined for each node \cite{Rey:2010ry}. 

In Section 2, after introducing the theories we will consider, we derive the perturbative series of the planar free energy, to all orders in the 't Hooft couplings $\lambda_i$. Let's present here the answer for the $\widehat{A_1}$ theory. It is convenient to define ${\cal F}_0(\lambda_1,\lambda_2)=F_0(\lambda_1,\lambda_2)-F_0(\lambda_1)^{{\cal N}=4}-F_0(\lambda_2)^{{\cal N}=4}$. The perturbative series is given by a sum over tree graphs,
\be
\begin{split}
{\cal F}_0(\lambda_1,\lambda_2) =\sum_{m=1}^\infty (-2)^m \sum_{n_1,\dots,n_m=2}^\infty
\frac{\zeta(2n_1-1)\dots \zeta(2n_m-1)}{n_1\dots n_m} (-1)^{n_1 + \dots + n_m} \\
\sum_{k_1=1}^{n_1-1} {2n_1\choose 2k_1} \dots \sum_{k_m=1}^{n_m-1} {2n_m \choose 2k_m}  
\sum_{\substack{\text{unlabeled trees}\\ \text{with $m$ edges}}} \frac{1}{|\text{Aut(T)}|}
\prod_{i=1}^{m+1} \bar {\cal V}_i \, ,
\end{split}
\label{planarfintro}
\ee
where the product at the end of the last line runs over the vertices of a tree, and $\bar {\cal V}_i$ are factors to be defined below. This expression is formally identical to the one found for ${\cal N}=2$ SQCD in \cite{Fiol:2020bhf}, except for the fact that now the factors $\bar {\cal V}_i$ depend on two 't Hooft couplings, $\lambda_1$ and $\lambda_2$. The terms in (\ref{planarfintro}) with a single value of the $\zeta$ function have already appeared in \cite{Pini:2017ouj}. In the perturbative expansion of ${\cal F}_0(\lambda_1,\lambda_2)$ above, each product of values of the $\zeta$ function is accompanied by a polynomial in $\lambda_1$ and $\lambda_2$, that can be rewritten as a palindromic polynomial in $\lambda_2/\lambda_1$. Intriguingly, up to the order we have checked explicitly, all such polynomials have all roots on the unit circle of the complex $\lambda_2/\lambda_1$ plane. This is of course reminiscent of the seminal work by Lee and Yang \cite{Lee:1952ig} for the zeros of the partition function of the ferromagnetic Ising model on a graph. We are able to prove this property for all the terms in (\ref{planarfintro}) with a single value of $\zeta$, and formulate two conjectures for general trees.

In Section 3, we compute the planar limit of the expectation value of the $1/2$ BPS circular Wilson loop defined for the gauge group in one of the two nodes of the $\widehat{A_1}$ theory, and in the fundamental representation. The answer is now given as a sum over rooted trees. This Wilson loop is defined for one of the two nodes of the quiver, so it does not transform nicely under the $\bZ_2$ symmetry of the theory. For this reason we consider $\vev{W}_\pm =\vev{W_1}\pm \vev{W_2}$ (with the ${\cal N}=4$ results subtracted). For $\vev{W}_\pm$ we find again that, up to the orders we have checked explicitly, all the polynomials in $\lambda_2/\lambda_1$ that appear have all roots on the unit circle.

In the appendices, we write the first terms in the explicit expansion of the planar free energy and expectation value of various Wilson loop operators.

This work leaves open a number of interesting problems. First, there are general arguments that the perturbative series of the planar limit of quantum field theories have finite radius of convergence \cite{Koplik:1977pf}. We have been able to determine the domain of convergence of just a small subset of the perturbative series found in this paper  - see also \cite{Pini:2017ouj} - but rigorously determining the full domain of convergence of the full perturbative series seems like a much harder problem. Second, in the main text we formulate two conjectures on the zeros of the polynomials that appear in the perturbative series of the planar free energy and expectation values of Wilson loops. It would be interesting to prove these conjectures, and further investigate if this property is related to the integrability of the planar limit of these theories \cite{Pomoni:2013po,Gadde:2010zi,Gadde:2012rv,Pomoni:2019oib}.

\section{The partition function of $\mathcal{N}=2$ quiver CFT}
In this section we introduce the theories we are going to study, and recall how supersymmetric localization reduces the evaluation of selected quantities to matrix integrals. In particular, we will study first the planar free energy of the theory. Following \cite{Billo:2017glv,Billo:2018oog,Billo:2019fbi}, the integrals are performed over the full Lie algebra instead of restricting to a Cartan subalgebra, and the 1-loop factor is rewritten as an effective action. We will focus on the planar limit and in this limit, as in \cite{Fiol:2020bhf},  we will unravel the underlying graph structure of the perturbative expansion.

Let us start by briefly reviewing the classification and field content of ${\cal N}=2$ superconformal quiver gauge theories with SU($N$) gauge groups. They are in one-to-one correspondence with simply-laced affine Lie algebras $\widehat{ADE}$, and thus follow an ADE classification \cite{Katz:1997eq}. The gauge sector and matter content are encoded in the extended Cartan matrix of the affine Lie algebra. The gauge group is
\be
\prod_i \text{SU}(n_iN) \, ,
\ee
where $n_i$ is the Dynkin index of the $i$-th node of the affine Dynkin diagram. The hypermultiplets transform in the representations
\be
\oplus a_{ij} \left(n_i N, \overline{ n_j N}\right) \, ,
\ee
where $a_{ij}$ is the adjacency matrix of the Dynkin diagram. 

These theories have a marginal coupling for each gauge group and, in the particular case where the complexified couplings satisfy 
\be
\tau_i = n_i \tau \,,
\ee
the quiver theory can be obtained as an orbifold of ${\cal N}=4$ SU($N$) super Yang-Mills by the discrete subgroup $\Gamma$ of SU(2) \cite{Katz:1997eq}, which also follow an ADE classification. These theories  can be engineered in string theory via a suitable brane configuration and even more, in a suitable limit, they admit a weakly curved gravity dual in terms of the $AdS_5 \times S^5  /\Gamma$ geometry  \cite{Kachru:1998ys, Lawrence:1998ja}. On the other hand, when all the couplings are set to zero except one, say $g_1$, the quiver theory reduces to  $\mathcal{N}=2$ SQCD.

After having reviewed ${\cal N}=2$ superconformal quiver theories, let's discuss supersymmetric localization for them. Following \cite{Pestun:2007rz} it is possible to localize the $\widehat{ADE}$ theories on $S^4$. It is also possible to localize the theory on a squashed sphere of parameter $b$ for which in the limit $b \to 1$ we recover the sphere, in such configuration the exact partition function is given by

\begin{equation}
	Z = \int \textnormal{d}a_I  \, \mathcal{Z}_{\text{1-loop}}(a_I,b) \lvert \, \mathcal{Z}_{\text{inst}}(a_I,b) \lvert^2 e^{-\sum_{I=1}^n \frac{8\pi^2}{g_I^2} \textnormal{Tr} a_I ^2} \, ,
	\label{def:part_funct_b}
\end{equation} 
where $a_I$ denotes the eigenvalues of the vector-multiplet scalars $\Phi_I$ restricted to the constant mode on $S^4$. In what follows we will be mostly interested in quantities that are relevant in the $b \simeq 1$ limit, such as the Wilson loop operator, or even more just observables defined on the sphere. As usual we will restrict our analysis to the zero-instanton sector, thus neglecting $\lvert \mathcal{Z}_{\text{inst}}\lvert^2$, and expanding (\ref{def:part_funct_b}) in b we obtain 
	\begin{equation}
		Z = \int \, \textnormal{d}a_I \, \mathcal{Z}_{\text{1-loop}}(a_I) \,  e^{-\sum_{I=1}^n \frac{8\pi^2}{g_I^2} \textnormal{Tr} a_I ^2} + \mathcal{O}((b-1)^2 ) \, ,
	\end{equation}
higher order terms in $b$ were studied before in \cite{Mitev:2014yba} and we refer the reader there for more details. 
The factor $\mathcal{Z}_{\textnormal{1-loop}}$ is the 1-loop contribution determined by the matter content. For instance for the $\widehat{A_{n-1}}$ theory it is given by
	\begin{equation}
		\mathcal{Z}_{\textnormal{1-loop}} = \prod_{I=1}^n \frac{\prod_{i<j}H^2(i a_i^I-i a_j^I)}{\prod_{i,j}H(i a_i^I-i a_j^{I+1})}	\, ,
	\end{equation}
where we identify the node $n+1$ with the first one and 
% The field content of the SU($N$)$\times$SU($N$) quiver consists of two vector multiplets in the adjoint: $(A_{\mu}^{(i)}, \Phi^{(i)}, \Phi^{'(i)})$, $i=1,2$, and bi-fundamental matter: $(X,Y,X^{\dagger},Y^{\dagger}): \, D_\mu X = \partial_\mu X + A_{\mu}^{(1)}X-X A_{\mu}^{(2)}$. The theory can be localized upon placing it on the four-sphere \cite{Pestun:2007rz} and as usual this reduces the path integral to a matrix model given by
% 	\begin{equation}
% 		Z = \int da db \frac{\Pi_{i<j}H^2(i a_i-i a_j) H^2(i b_i-i b_j)}{\Pi_{i,j} H^2(i a_i-i b_j)} e^{-\frac{8\pi^2}{g_1^2} \textnormal{Tr}\,{a^2}-\frac{8\pi^2}{g_2^2} \textnormal{Tr}\, b^2}
% 		\label{def:mmodel}
% 	\end{equation}
% where we have set ${\cal Z}_{\text{ins}}=1$. The integration variables $a_i$, $b_i$ are the eigenvalues of the vector-multiplet scalars $\Phi^{(i)}$, with $i=1,2$ respectively, restricted to the constant mode on $S^4$. 
$H(x)$ is the Barnes function whose expansion is given by
	\begin{equation}
		\log H(x) = -(1+\gamma) x^2 - \sum_{n=2}^{\infty} \frac{\zeta(2n-1)}{n} x^{2n} \, .
		\label{def:barnes}
	\end{equation}

Following the previous works \cite{Billo:2017glv,Billo:2018oog,Billo:2019fbi} the strategy will be once again to interpret the matter content as an effective action 
	\begin{equation}
		S_{int} = - \log \mathcal{Z}_{\text{1-loop}} \, .
	\end{equation}
Given that the theory is conformal for arbitrary values of the couplings, the quadratic terms in (\ref{def:barnes}) will exactly cancel and the effective action will start at order $g_i^4$. 

Let us first illustrate the process with the $\widehat{A_1}$ quiver since the extension to the general case is straightforward. In this case the field content of the $\widehat{A_1}$ quiver consists of two vector multiplets in the adjoint: $(A_{\mu}^{I}, \Phi^{I}, \Phi^{'I})$, $I=1,2$, and bi-fundamental matter: $(X,Y,X^{\dagger},Y^{\dagger}): \, D_\mu X = \partial_\mu X + A_{\mu}^{1}X-X A_{\mu}^{2}$. The l-loop factor reduces to %Upon renaming the vevs $a_1 \to a$, $a_2 \to b$ the 1-loop factor reduces to 
	\begin{equation}
		\mathcal{Z}_{\textnormal{1-loop}} = \frac{\prod_{i<j}H^2(i a_i^1 - i a_j^1) H^2(i a_i^2 - i a_j^2)}{\prod_{i,j} H^2(i a_i^1 - i a_j^2)} \, .
	\end{equation}
Following the procedure presented in \cite{Fiol:2020bhf} and using (\ref{def:barnes}) it is possible to arrive to the effective action, obtaining 
	\begin{equation}
	\begin{split}
		S_{int} = \sum_{n=2}^{\infty} &\frac{\zeta(2n-1) (-1)^n }{n} \left[ \sum_{k=1}^{n-1} {2n \choose 2k} \left(\textnormal{Tr} \,a_1^{2(n-k)} \textnormal{Tr} \,a_1^{2k}+ \textnormal{Tr} \,a_2^{2(n-k)} \textnormal{Tr} \,a_2^{2k} - 2 \textnormal{Tr} \,a_1^{2(n-k)} \textnormal{Tr} \,a_2^{2k} \right)\right. \\
		& \left. - \sum_{k=1}^{n-2} {2n \choose 2k+1} \left( \textnormal{Tr} \,a_1^{2(n-k)-1} \textnormal{Tr} \,a_1^{2k+1}+ \textnormal{Tr} \,a_2^{2(n-k)-1} \textnormal{Tr} \,a_2^{2k+1}- 2 \textnormal{Tr} \,a_1^{2(n-k)-1} \textnormal{Tr} \,a_2^{2k+1} \right) \right] \, ,
	\end{split}
	\label{eq:eff_action}
	\end{equation}
where all traces are in the fundamental representation of the respective gauge group. Let's comment upon a couple of features of this result: first, as we already encountered in our previous work for theories with simple gauge groups \cite{Fiol:2020bhf}, the effective action involves infinite sums of double-trace terms, that split into even and odd powers. By the same large $N$ counting arguments as in \cite{Fiol:2020bhf},  the odd powers will not contribute to the planar computations, so we discard such terms in what follows. Second, the pattern of double-trace terms in (\ref{eq:eff_action}) is dictated by the Cartan matrix of $\widehat{A_1}$,
\be
\frac{1}{2}\, C=
\begin{pmatrix}
1 & -1 \\
-1 & 1
\end{pmatrix} \, .
\label{cartanau}
\ee
This shouldn't be a surprise, since for ${\cal N}=2$ quiver superconformal field theories, the matter content is fixed by the 1-loop $\beta$ functions, which are captured by the generalized Cartan matrix \cite{Katz:1997eq}. This last observation allows us to generalize (\ref{eq:eff_action})  to arbitrary ${\cal N}=2$ superconformal quiver theory. The effective action, keeping just the terms with even powers, is
\begin{equation}
%		\begin{split}
%			S_{\textnormal{int}} = \sum_I \sum_{n=2}^\infty \frac{\zeta(2n-1)(-1)^n}{n} \sum_{k=1}^{n-1} {2n \choose 2k} \left( \textnormal{Tr}a_I^{2(n-k)}\textnormal{Tr}%a_I^{2k} - \textnormal{Tr}a_I^{2(n-k)}\textnormal{Tr}a^{2k}_{I+1} \right) =	\\
S_{int} = \frac{1}{2} \sum_{I,J} C_{IJ} \sum_{n=2}^\infty \frac{\zeta(2n-1)(-1)^n}{n} \sum_{k=1}^{n-1} {2n \choose 2k} \textnormal{Tr}a_I^{2(n-k)}\textnormal{Tr}a_J^{2k} \, ,
	\end{equation}
where $C_{IJ}$ is the Cartan matrix of the corresponding affine Lie algebra.

\subsection{Planar free energy}
We turn now to the large $N$ limit of the free energy on $S^4$, $F(\lambda_i, N) = \log Z_{S^4}$. In fact, as usual, we will compute the difference of free energy with the Gaussian model, $ {\cal F}(\lambda_i,N)\equiv F(\lambda_i,N)-\sum_i F(\lambda_i)^{{\cal N}=4}$. Our goal is to determine the leading term in the large $N$ expansion, \emph{i.e.} $F(\lambda_i, N) = F_0(\lambda_i) N^2 + \cdots$. In general we have

\begin{equation}
	F(\lambda_i, N) = \log Z_{S^4} = \sum_{m=1}^\infty \frac{(-1)^{m+1}}{m} \left( \sum_{k=1}^{\infty} \frac{(-1)^k}{k!} \langle S_{int}^k \rangle \right)^m \, ,
	\label{eq:free_exp}
\end{equation} 
the free energy scales like $N^2$ in the planar limit, so there are many cancellations in (\ref{eq:free_exp}) and we need to fully identify the $N^2$ terms from (\ref{eq:free_exp}) that survive these cancellations. The argument to extract those terms is exactly the same as in our recent work \cite{Fiol:2020bhf}: for a disconnected $2m$-point function, the pieces that scale like $N^2$ are products of $m+1$ connected correlators.  These connected correlators in the planar limit are given by \cite{tutte} (see also \cite{Gopakumar:2012ny} for a more recent derivation) 
	\begin{equation}
		\vev{\traza{2k_1} \traza{2k_2} \dots \traza{2k_n}}_c=   {\cal V}(k_1,\dots,k_n)   \tilde{\lambda}^d N^{2-n} , \quad \tilde{\lambda} = \frac{\lambda}{16\pi^2},
		\label{gopakumarf}
	\end{equation}
with
\be 
{\cal V}(k_1,\dots,k_n)= \frac{(d-1)!}{(d-n+2)!} 	\prod _{i=1}^n \frac{(2k_i)!}{(k_i-1)! k_i!} , \quad d=\sum_{i=1}^n k_i .
\label{gpfactor}
\ee
The products of $m+1$ connected correlators that contribute to the planar free energy are those where the $2m$ traces are distributed in a way that can be characterized by a tree graph \cite{Fiol:2020bhf}: for each correlator introduce a vertex, and join them by an edge if they have operators from the same double-trace. The contributions to $\mathcal{F}_0(\lambda)$ at fixed order in the number of values of $\zeta$ function are then obtained following a similar procedure as in our recent work \cite{Fiol:2020bhf}, but with a couple of modifications. Terms with $m$ values of the $\zeta$ function have $m$ pairs of traces, coming from $m$ double-trace terms, which are of the form $C_{IJ}  \textnormal{Tr}\,a_I^{2(n-k)}\textnormal{Tr}\,a_J^{2k} $. 

To find the contribution to the planar free energy at this order, first draw all the trees with $m$ edges. For every tree, assign each of the $m$ double-traces to one of the $m$ edges; this labels the $m$ edges of the tree, turning it into a edge-labeled tree. Next, add an arrow to each of the $m$ edges, turning the tree into a directed edge-labeled tree. Assign $\textnormal{Tr}\,a_I^{2(n-k)}$ to the vertex at the start ({\it i.e.} origin of the arrow) of the $i$-th edge. Assign $\textnormal{Tr}\,a_J^{2k} $ to the vertex at the end ({\it i.e.} end of the arrow) of the $i$-th edge. This procedure assigns to each of the $m+1$ vertices a number of traces equal to its degree $\alpha_j$, {\it i.e.}  the number of edges connected to that vertex. For each vertex, consider now the connected correlator of all its trace operators and assign it its numerical factor ${\cal V}_j$, eq. (\ref{gpfactor}), times $\tlambda_j^{d_j}$, with $j=1,\dots,m+1$. For the connected correlator to be nonzero, all traces at  a given vertex must be of the same matrix, and this enforces that they have the same index. Finally, multiply the contribution of this tree graph by a product of $m$ components of the Cartan matrix, one per edge, with the indices fixed by those at the vertices of each edge. Summing over all the possible choices, we arrive at

\be
\begin{split}
{\cal F}_0(\lambda_1,\dots ,\lambda_n) =\sum_{m=1}^\infty \frac{(-1)^m }{m!}\sum_{n_1,\dots,n_m=2}^\infty
\frac{\zeta(2n_1-1)\dots \zeta(2n_m-1)}{n_1\dots n_m} (-1)^{n_1 + \dots + n_m} \\
\sum_{k_1=1}^{n_1-1} {2n_1\choose 2k_1} \dots \sum_{k_m=1}^{n_m-1} {2n_m \choose 2k_m}  
\frac{1}{2^m} \sum_{\substack{\text{ directed trees}\\ \text{with m labeled edges}}} \sum_{I,J} C_{I_1 J_1} \dots C_{I_m J_m} 
\prod_{i=1}^{m+1} \tlambda_{I_i}^{d_i} {\cal V}_i \, .
\end{split}
\label{planarfade}
\ee
This expression is the perturbative series for the planar free energy of any ${\cal N}=2$ superconformal quiver theory, with quiver determined by the affine Lie algebra with Cartan matrix $C$. In what follows, we will discuss mostly the simplest quiver theory, $\widehat{A_1}$, that has gauge group SU($N$)$\times$SU($N$), and Cartan matrix (\ref{cartanau}). This means that double-traces where both operators belong to the same gauge group, {\it e.g. } $\textnormal{Tr}\,a_1^{2(n-k)}\textnormal{Tr}\,a_1^{2k}$   are weighted with a $+1$, while mixed double-traces, {\em e.g. } $\textnormal{Tr}\,a_1^{2(n-k)}\textnormal{Tr}\,a_2^{2k}$ are weighted with a $-1$. The overall sign of a given product of correlators is then $-1$ raised to the number of mixed double-traces. These signs can be transferred from the edges to the vertices: just assign an extra factor $(-1)^{\alpha_j}$ to all vertices of the tree corresponding to correlators of, say, the second gauge group (this choice is arbitrary and the final result is independent of it). To convince oneself that these two rules are the same, write every sign on top of the edges of the tree: if it is a $-1$ assign it to the vertex with operators of the second gauge group. If it is a $+1$, and it is joining two vertices with operators of the second gauge group, just write $+1=(-1)(-1)$ and again assign one $-1$ to each vertex. Then each vertex contributes a factor
\be
\bar {\cal V}(x_1,\dots,x_\alpha)={\cal V}(x_1,\dots,x_\alpha) \left( \tlambda_1^{\sum_i x_i} +(-1)^\alpha \tlambda_2^{\sum_i x_i}\right) \, ,
\label{modgpfactor}
\ee
and the generic expression (\ref{planarfade}) simplifies to
\be
\begin{split}
{\cal F}_0(\lambda_1,\lambda_2) =\sum_{m=1}^\infty \frac{(-1)^m }{m!}\sum_{n_1,\dots,n_m=2}^\infty
\frac{\zeta(2n_1-1)\dots \zeta(2n_m-1)}{n_1\dots n_m} (-1)^{n_1 + \dots + n_m}\\
\sum_{k_1=1}^{n_1-1} {2n_1\choose 2k_1} \dots \sum_{k_m=1}^{n_m-1} {2n_m \choose 2k_m}  
\sum_{\substack{\text{directed trees}\\ \text{with $m$ labeled edges}}} 
\prod_{i=1}^{m+1} \bar {\cal V}_i \, .
\end{split}
\label{planarfa1}
\ee
Finally, by exactly the same arguments as in our previous paper \cite{Fiol:2020bhf}, the last sum can be reduced to a sum over unlabeled trees
\be
\begin{split}
{\cal F}_0(\lambda_1,\lambda_2) =\sum_{m=1}^\infty (-2)^m \sum_{n_1,\dots,n_m=2}^\infty
\frac{\zeta(2n_1-1)\dots \zeta(2n_m-1)}{n_1\dots n_m} (-1)^{n_1 + \dots + n_m}\\
\sum_{k_1=1}^{n_1-1} {2n_1\choose 2k_1} \dots \sum_{k_m=1}^{n_m-1} {2n_m \choose 2k_m}  
\sum_{\substack{\text{unlabeled trees}\\ \text{with $m$ edges}}} \frac{1}{|\text{Aut(T)}|}
\prod_{i=1}^{m+1} \bar {\cal V}_i \, .
\end{split}
\label{planarf2}
\ee

Let's mention a further property of ${\cal F}_0(\lambda_1,\lambda_2)$. Since ${\cal F}_0(\lambda_2,\lambda_1)={\cal F}_0(\lambda_1,\lambda_2)$ and ${\cal F}_0(\lambda_1,\lambda_1)=0$, it follows that ${\cal F}_0(\lambda_1,\lambda_2)$  has a double zero,
\be
{\cal F}_0(\lambda_1,\lambda_2)=(\lambda_1-\lambda_2)^2 f(\lambda_1,\lambda_2) \, ,
\label{doublezero}
\ee
this implies that at the orbifold point $\lambda_1=\lambda_2$ not just the free energy, but also its first derivative with respect to $\lambda$ coincides with the ${\cal N}=4$ result. To see that this property is implied by our result (\ref{planarf2}), we are going to prove that the contribution of every tree to (\ref{planarf2}) is of the form 
\be
(\tlambda_1-\tlambda_2)^{v_{\text{odd}}} \, \, p(\tlambda_1,\tlambda_2) \, ,
\label{treepoly}
\ee
where $v_{\text{odd}}$ is the number of vertices of the tree with odd degree, and $p(\tlambda_1,\tlambda_2)$ is a symmetric polynomial in $\tlambda_1$ and $\tlambda_2$ with positive coefficients. This follows from inspection of the factor attached to each vertex, (\ref{modgpfactor}). When the degree $\alpha$ of a vertex is odd, $\tlambda_1 =\tlambda_2$ is a simple root of that factor. After pulling out these factors, what is left is a polynomial with positive coefficients. As a check, notice that $v_{\text{odd}}$ is always even: for a tree with $m+1$ vertices, $\sum_{i=1}^{m+1} \alpha_i=2m$, and since $\sum_i \alpha_i^{\text{even}}$ is even, $\sum_i \alpha_i^{\text{odd}}$ must be even also, which implies that $v_{\text{odd}}$ is even. This concludes the argument for (\ref{treepoly}). Now, since every tree has at least two vertices of degree one, $v_{\text{odd}} \geq 2$, and (\ref{doublezero}) follows.

To illustrate (\ref{planarf2}), let's work out the first terms. The $m=1$ terms in (\ref{planarf2}) are terms with a single value of $\zeta$ \cite{Pini:2017ouj}. To write them, it is convenient to first recall the definition of the Narayana numbers
\be
N(n,k)=\frac{1}{n}{n \choose k} {n \choose k-1} \ ,
\ee
and the Narayana polynomials
\be
\mathfrak{C}_n(t)=\sum_{k=0}^{n-1} N(n,k+1) t^k \, ,
\label{narapol}
\ee
that satisfy $\mathfrak{C}_n(1)=\mathcal{C}_n$ with ${\cal C}_n$ the Catalan numbers. At this order, we have to consider  trees with two vertices. There is just one such tree, and both vertices have degree one. Then,
\begin{multline}
{\cal F}_0(\lambda_1,\lambda_2) \vert_\zeta = - \sum_{n=2}^\infty \frac{\zeta(2n-1)}{n} (-1)^n \sum_{k=1}^{n-1} {2n \choose 2k} \mathcal{C}_{n-k} \mathcal{C}_k \left( \tilde{\lambda}_1^{n-k} -\tilde{\lambda}_2^{n-k} \right) \left( \tilde{\lambda}_1^k-\tilde{\lambda}_2^k  \right) \\
=  - \sum_{n=2}^\infty \frac{\zeta(2n-1)}{n} (-1)^n \mathcal{C}_n \tlambda_1^n\left[ \left(1 + \frac{\tlambda_2^n}{\tlambda_1^n}\right)\mathcal{C}_{n+1} - 2 \mathfrak{C}_{n+1}\left(\frac{\tlambda_2}{\tlambda_1}\right) \right] \, ,
\label{linearfree}
\end{multline}
where to avoid confusion, the first term in the parenthesis involves the Catalan number ${\cal C}_{n+1}$, and the second one the Narayana polynomial $\mathfrak{C}_{n+1}(\tlambda_2/\tlambda_1)$. A first question we can ask about this series is what is its domain of convergence in $\bC^2$. As pointed out in \cite{Pini:2017ouj, Fiol:2020bhf}, when $\lambda_2=0$ it is straightforward to prove that the radius of convergence is $\lambda_1=\pi^2$, and the same holds, {\em mutatis mutandi}, when $\lambda_1=0$. When both couplings are different from zero, since ${\cal F}_0(\lambda_1,\lambda_1)=0$ the series trivially converges when both couplings are equal. When the two couplings are different, one of them is larger, say $\lambda_1$, applying the quotient criterion it follows that for any $\vert\lambda_2\vert<\vert\lambda_1\vert\leq \pi^2$, the series is convergent. All in all, this series is convergent in $|\lambda_1|\leq \pi^2$, $ |\lambda_2|\leq \pi^2$ plus the $\lambda_1=\lambda_2$ line. 

For ${\cal N}=2$ superconformal field theories with a simple gauge group, terms with a fixed number of values of the $\zeta$ function form an infinite series. In \cite{Fiol:2020bhf} we sketched an argument that all these series have the same radius of convergence. It seems possible that this property extends to quiver theories.

Let's work out a couple more of terms in (\ref{planarf2}). Terms with two values of the $\zeta$ function are given by a sum over trees with two edges. There is just one tree with two edges, and its vertices have degrees $(1,2,1)$. As a last example, terms with three values of the $\zeta$ function are given by a sum over trees with three edges. There are two such unlabeled trees. The degrees are $(1,2,2,1)$ for the first tree, and $(3,1,1,1)$ for the second, all these trees are despicted in fig. (\ref{img:treesb2}) and (\ref{img:treesb3}). Up to this order,

\begin{equation}
\begin{split}
{\cal F}_0(\lambda_1,\lambda_2) = &  -  \sum_{n=2}^\infty \frac{\zeta(2n-1)}{n} (-1)^n \sum_{k=1}^{n-1} {2n \choose 2k} \mathcal{V}(n-k) \mathcal{V}(k)  (\tlambda_1^{n-k}-\tlambda_2^{n-k})
(\tlambda_1^k-\tlambda_2^k) \\
& +\frac{1}{2} \sum_{n_i=2}^\infty \frac{\zeta(2n_i-1)}{n_1 n_2} (-1)^{n_1 + n_2} \sum_{k_i=1}^{n_i-1} {2n_i \choose 2k_i }  4 {\cal V}(k_1){\cal V}(n_1-k_1,n_2-k_2){\cal V}(k_2) \\ 
&  (\tlambda_1^{k_1}-\tlambda_2^{k_1}) (\tlambda_1^{n_1-k_1+n_2-k_2}+\tlambda_2^{n_1-k_1+n_2-k_2})(\tlambda_1^{k_2}-\tlambda_2^{k_2}) \\
& -\frac{1}{3!} \sum_{n_i=2}^\infty \frac{\zeta(2n_i-1)}{n_1 n_2 n_3} (-1)^{n_1 + n_2 + n_3} \sum_{k_i=1}^{n_i-1} {2n_i \choose 2k_i} 8 \Bigl[ 3 {\cal V}(n_1-k_1) {\cal V}(k_1,n_2-k_2) \\ & \times {\cal V}(k_2,n_3-k_3)  {\cal V}(k_3) (\tlambda_1^{n_1-k_1}-\tlambda_2^{n_1-k_1}) (\tlambda_1^{k_1+n_2-k_2}+\tlambda_2^{k_1+n_2-k_2}) \\
& \times (\tlambda_1^{k_2+n_3-k_3}+\tlambda_2^{k_2+n_3-k_3}) (\tlambda_1^{k_3}-\tlambda_2^{k_3}) + {\cal V}(n_1-k_1,n_2-k_2,n_3-k_3) {\cal V}(k_1) {\cal V}(k_2) {\cal V}(k_3) \\
& \left. \times (\tlambda_1^{n_1-k_1+n_2-k_2+n_3-k_3}-\tlambda_2^{n_1-k_1+n_2-k_2+n_3-k_3}) (\tlambda_1^{k_1}-\tlambda_2^{k_1}) (\tlambda_1^{k_2}-\tlambda_2^{k_2}) (\tlambda_1^{k_3}-\tlambda_2^{k_3}) \right] +{\cal O}(\zeta^4) \, .
 \end{split}
 \label{res:FO}
\end{equation}
As a first check, when either of the two couplings vanishes, we recover the result of ${\cal N}=2$ SCQD presented in \cite{Fiol:2020bhf}. Also, in this expression we can see rather explicitly that at every order the contribution has at least a double zero $(\lambda_1-\lambda_2)^2$. In Appendix A we have written the outcome of these sums, up to order $\lambda^6$.

\subsection{The Lee-Yang property of the planar free energy expansion.}

\begin{figure}[h]
		\includegraphics[scale=1]{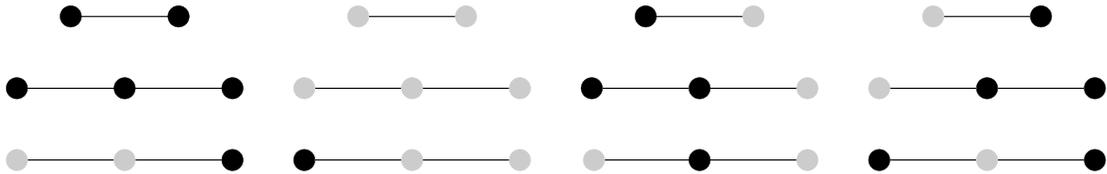}
		%\put(-380,-15){(a)}
		%\put(-271,-15){(b)}
		%\put(-161,-15){(c)}
		%\put(-53.5,-15){(d)}
		\caption{Trees contributing to the first and second order expansion of the free energy.} %different colors correspond to different gauge groups from SU($N$)$\times$SU($N$). (a) \& (b) are trees arising from contractions within the same group, we call them positive/negative. (c) \& (d) are trees arising from the interaction term in (\ref{eq:eff_action}). }
		\label{img:treesb2}
	\end{figure}

\begin{figure}[h]
	\begin{center}
		\includegraphics[scale=1]{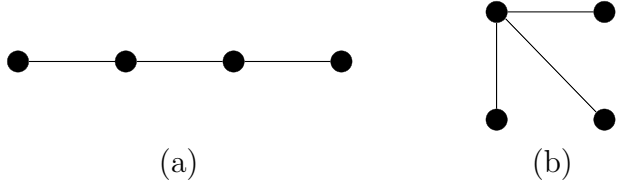}
		\put(-173,-15){(a)}
		\put(-32,-15){(b)}
		\caption{The two trees with three edges: $(a)$ Tree with vertices of degrees (1,2,2,1). $(b)$ Tree with vertices of degrees (3,1,1,1). There are 16 ways to color each of them.}%Trees arising from ${\cal V}(n_1-k_1,n_2-k_2){\cal V}(n_3-k_3,k_1){\cal V}(k_2){\cal V}(k_3)$. $(b)$ Trees arising from ${\cal V}(n_1-k_1,n_2-k_2,n_3-k_3){\cal V}(k_1){\cal V}(k_2){\cal V}(k_3)$}
		\label{img:treesb3}
	\end{center}
	\end{figure}

We would like to discuss one further property of the perturbative expansion (\ref{planarf2}). Notice that the contribution of a given tree is obtained by summing over all the possible ways to assign one gauge group, $1$ or $2$, to each vertex in the tree, see figures (\ref{img:treesb2}) and (\ref{img:treesb3}). This is reminiscent of the Ising model defined on that tree, where on each vertex we can have a spin up or down. It is indeed possible to construct a generalized Ising-type model, with inhomogeneous external magnetic field, whose partition function yields each tree contribution in (\ref{planarf2}). This generalized Ising model is admittedly a bit contrived, but following the classical work by Lee and Yang \cite{Lee:1952ig}, it motivates the study of the zeros of its partition function. 

In more detail, every tree graph contributes to the planar energy in (\ref{planarf2}) a homogeneous polynomial in $\lambda_1$ and $\lambda_2$. Being homogeneous, these polynomials can be thought of as polynomials of a single variable $\lambda_2/\lambda_1$. Inspired by the classical work by Lee and Yang \cite{Lee:1952ig} on the ferromagnetic Ising model, we are going to put forward two conjectures regarding the zeros of these polynomials: first, that for a given tree, all the zeros of the corresponding polynomial are on the unit circle in the complex $\lambda_2/\lambda_1$ plane. Second, that when we sum the contributions from different trees with the same number of nodes, the same property holds. 

To provide context, let's start by briefly recalling the definition of the Ising model on a graph and the Lee-Yang theorem. Let $G$ be a finite graph, $E$ its set of edges and $V$ its set of vertices. The Ising model on $G$ is defined by assigning to each vertex $i\in V$, a $\sigma_i=\pm 1$ (spin up/down). The Hamiltonian is
\be
{\cal H}= -J \sum_{i-j \in E} \sigma_i \sigma_j - H \sum_{i\in V} \sigma_i \, ,
\ee
with $J$ the coupling among spins and $H$ the external magnetic field. The partition function can be written as
\be
Z(\beta J,\beta H)=\sum_{\text{all states}} e^{-\beta {\cal H}} \,= \, e^{\beta J |E|-\beta H |V|} \sum_{\text{all states}} e^{-2\beta J e_\pm} \, \, e^{2\beta H v_\uparrow} \, ,
\ee
where $e_\pm$ is the number of edges connecting different spins, and $v_\uparrow$ the number of spins up in a given configuration. Define $\tau =e^{-2\beta J}$, $x=e^{2\beta H}$. The last sum defines a polynomial palindromic in $x$,
\be
P(\tau,x)=\sum_{\text{all states}} \tau^{e_\pm} \, x^{v_\uparrow} \, .
\ee

In  \cite{Lee:1952ig}, Lee and Yang proved that for $\tau\in [-1,1]$, the  polynomials $P(\tau,x)$ have all their $x$ roots on the unit circle. In fact, they proved it for arbitrary ferromagnetic couplings $J_{ij}\geq 0$, and different magnetic fields per site $H_i$.

To construct an Ising-type model whose partition function yields the polynomials that appear in  (\ref{planarf2}), proceed as follows. Take the graph G to be a tree T, \begin{enumerate}
\item{Assign a positive integer $n_i$ to each of the $e$ edges of the tree graph.}
\item{For every edge,  split $n_i$ into two positive integers, $n_i=k_i+(n_i-k_i)$ and assign each of these two integers to one of the vertices at the ends of that edge.}
\item{ Then, if a vertex has degree $d_j$ this procedure assigns to that vertex $d_j$ integers. Let $m_j$ be the sum of these integers at a given vertex; the magnetic field at that vertex is then $m_j H$.}
\end{enumerate}
So far, for a fixed partition of all $n_i$, this is a peculiar way to assign external magnetic fields that are different at each vertex. This defines
\be
P(\tau,x,k_i,n_i)= \sum_{\text{all states}}  \tau^{e_\pm} \, \prod_{\substack{\text{vertices} \\ \text{with spin up}}} x^{m_j} \, ,
\ee
Lee and Yang already proved (lemma in Appendix II of \cite{Lee:1952ig})  that all the zeros of these polynomials are on the unit circle. Finally, consider the sum over all the partitions of each of the $n_i$ into two 
\be
P(\tau,x, n_1,\dots, n_e)= \sum_{k_1=1}^{n_1-1}\dots \sum _{k_e=1}^{n_e-1}  \rho(k_i,n_i) \, \sum_{\text{all states}}  \tau^{e_\pm} \, \prod_{\substack{\text{vertices} \\ \text{with spin up}}} x^{m_j} \, ,
\label{polbeforesum}
\ee
where $\rho(k_i,n_i)$ is a distribution that weights different configurations. The contribution of every tree to the planar free energy in (\ref{planarf2}) is obtained from the free energy of this Ising-type model, by setting $\tau=-1$, $x=\lambda_2/\lambda_1$, and the distribution 
\be
\rho(k_i,n_i)= {2n_1 \choose 2k_1}\dots {2n_m \choose 2k_m} \prod_{i=1}^m{\cal V}_i \, .
\ee
The main reason we have defined this family of Ising-type models is that there is numerical evidence that suggests that they share the Lee-Yang property with the original Ising model. This leads us to formulate the following two conjectures:

\underline{Conjecture 1}: For any tree with $e$ edges, any fixed positive integers $n_1,\dots,n_e$ and arbitrary $\rho(k_i,n_i)>0$ the polynomials $P(\tau,x, n_1,\dots, n_e)$ have all their $x$ roots on the unit circle.

\underline{Conjecture 2}: If we sum the polynomials of all the trees with the same number of edges, the resulting polynomial still has the Lee-Yang property.

We can prove the first conjecture in the particular case of the simplest tree. In this case, (\ref{polbeforesum}) is simply
\be
P(\tau,x,k,n)=x^n+\tau x^{n-k}+\tau x^k +1 \, ,
\ee
that for $|\tau| \leq 1$ has its roots on the unit circle. Then 
\be
P(\tau,x,k,n)=\sum_{k=1}^{n-1} \rho(n,k) \left(x^n+\tau x^{n-k}+\tau x^k +1\right) \, ,
\ee
with arbitrary $\rho(n,k)>0$. To prove that these polynomials have their roots on the unit circle, we make use of the following theorem  \cite{lakatos}: if $P(x) =A_nx^n + A_{n-1}x^{n-1}+\dots +A_1x+A_0$ is a palindromic polynomial and $2|A_n| \geq \sum_{j=1}^{n-1} |A_j|$, then all its zeros are in the unit circle. In our case, the inequality in the theorem is satistifed as long as $|\tau|\leq 1$, so the result follows. Back to the free energy of the quiver theory, one can check indeed that the polynomials in  the expansion (\ref{linearfree}) have the Lee-Yang property.

We haven't been able to prove these two conjectures for arbitrary tree graphs. After the seminal work \cite{Lee:1952ig}, the proof of the Lee-Yang unit circle theorem has been extended to many other systems, see {\it e.g. }\cite{asano, ruelle}. It would be interesting to see if any of these arguments can be adapted to prove our conjectures.

%{\tiny 
%From the figure (\ref{img:treesb2}) and (\ref{res:FO}) we can understand the underlaying structure of the free energy, in $(a)$, $(b)$ we recover the result presented in  \cite{Fiol:2020bhf}  for $\mathcal{N}=2$ up to order $\beta^2$. We see that the structure of the expansion corresponds to the sum of the result but now with two different coupling constants (one for each SU($N$))\footnote{This can be easily generalized to more groups.}. On the other hand the trees depicted in $(b)$, $(c)$ correspond to new structures arising from the interaction term between the gauge groups in (\ref{eq:eff_action}), we see that the ones in $(d)$ are the \emph{negatives} of the ones in $(c)$. }

\section{Wilson loop in the large $N$ limit}
For each of the gauge groups of the quiver theory, we can define a 1/2 BPS Wilson loop, with circular contour in Euclidean signature. The evaluation of its expectation value reduces to a matrix integral thanks to supersymmetric localization. We will now evaluate the planar limit of this expectation value and show that the perturbative series involves a sum over rooted trees. While the Wilson loop can be defined for arbitrary representations of the gauge group, in order to take advantage of the results of \cite{Gopakumar:2012ny, Fiol:2020bhf}, we will restrict its study to the fundamental representation
	\begin{equation}
		\langle W^{I} \rangle = \langle \frac{1}{N} \textnormal{Tr}_F \mathcal{P} \exp \oint_{\mathcal{C}}ds \left( i A_\mu^{I}(x)\dot{x}^\mu + \Phi^{I}(x)\lvert\dot{x}\lvert \right) \rangle  \, ,
	\end{equation}
where $I =1,\cdots,n $. The theory can be localized \cite{Pestun:2007rz} on the sphere with squashing parameter $b$, where $b=1$ corresponds to $S^4$, in such case the vev of the $1/2$ BPS Wilson loop reduces to 
	\begin{equation}
		\langle W^{\pm}_{I} \rangle = \frac{1}{Z} \int da_I \textnormal{Tr}\left( e^{-2\pi b^{\pm} a_{I}} \right) e^{-\sum_{I=1}^n \frac{8\pi^2}{g_I^2} \textnormal{Tr} a_I ^2} \mathcal{Z}_{\textnormal{1-loop}}(a_I,b)\lvert \mathcal{Z}_{\text{inst}}(a_I,b)\lvert^2 \, ,
	\end{equation}
now $\pm$ represents the two different trajectories in which we can compute the Wilson loop on the squashed sphere \cite{Mitev:2015oty}; from now on we will avoid the $\pm$ to make the notation less cumbersome, bearing in mind that in order to switch between trajectories we need to make the replacement $b \to b^{-1}$ in the following results. Once again we will consider the 1-loop contribution as an effective action, given by (\ref{eq:eff_action}), and as discussed on the previous section we will compute the large $N$ limit of this interacting theory while restricting ourselves to the zero-instanton sector. We are interested in observables that are only sensitive to the linear dependence of $\vev{W_b}$ in $(b-1)$, and since the dependence of $\mathcal{Z}_{\textnormal{1-loop}}(a_I,b)$ is quadratic in $b-1$, for our purposes we can compute $\vev{W_b}$ directly on $S^4$ \cite{Fiol:2015spa},
\begin{equation}
		\langle W^{\pm}_{I} \rangle = \frac{1}{Z} \int da_I \textnormal{Tr}\left( e^{-2\pi b^{\pm} a_{I}} \right) e^{-\sum_{I=1}^n \frac{8\pi^2}{g_I^2} \textnormal{Tr} a_I ^2} \mathcal{Z}_{\textnormal{1-loop}}(a_I) + \mathcal{O}((b-1)^2) \, .
	\end{equation}

Let us expand the Wilson loop insertion
	\begin{equation}
		\langle W_{I}\rangle = \sum_{l=0}^\infty \frac{(4\pi^2 b^2)^l}{(2l)!} \frac{\langle N^{-1} \textnormal{Tr}\, a_{I}^{2l} e^{-S}\rangle}{\langle e^{-S} \rangle} \, .
	\end{equation}
As argued in our recent work \cite{Fiol:2020bhf}, the large $N$ expansion of this expectation value scales like $N^0$, so given the overall normalization factor $1/N$, the relevant terms to keep from $\vev{ \textnormal{Tr}\, a_I^{2l} S^m}$ are products of $m+1$ connected correlators. Now there are $2m+1$ traces to be distributed in $m+1$ correlators, but since  $\vev{ \textnormal{Tr}\, a_I^{2l} }$ can't be by itself, we effectively have to distribute $2m$ traces into the $m+1$ connected correlators, which is the by now familiar sign that the possibilities are given by tree graphs. As in \cite{Fiol:2020bhf}, one of the vertices is singled out by the presence of $\vev{ \textnormal{Tr}\, a_I^{2l} }$, so these are rooted trees. The correlator that contains $\vev{ \textnormal{Tr}\, a_I^{2l} }$ is a correlator of $a_I$ operators, so it involves the $\lambda_I$ coupling; by convention, the root vertex corresponding to this correlator will be referred as the vertex 1. The remaining $m$ correlators can be either products of $a_I$ traces or $a_J$ traces. As we found in the evaluation of the planar free energy in the previous section, this is accounted for by modifying the numerical factor of the correlator by a weighted sum over the coupling. eq. (\ref{modgpfactor}). All in all, for the case of $\widehat{A_1}$
\be
\begin{split}
\vev{W_1}-\vev{W_1}_0= \sum_{l=1} \frac{(2\pi b)^{2l}}{(2l)!}   \sum_{m=1}^\infty (-2)^m \sum_{n_1,\dots,n_m=2}^\infty \frac{\zeta(2n_1-1)\dots \zeta(2n_m-1)}{n_1\dots n_m} (-1)^{n_1 + \dots + n_m}\\
\sum_{k_1=1}^{n_1-1} {2n_1 \choose 2k_1} \dots \sum_{k_m=1}^{n_m-1} {2n_m \choose 2k_m} \sum_{\substack{\text{unlabeled rooted trees}\\ \text{with $m$ edges}}}
\frac{1}{|\text{Aut(T)}| } \tlambda_1^{d_1} {\cal V}_1 \prod_{i=2}^{m+1} \bar {\cal V}_i \, ,
\end{split}
\ee
In the language of Ising-type models on trees introduced in the previous section, we can think of the Wilson loop insertion as a spin that is pinned to be up, at the rooted vertex. To illustrate this result, let's expand it up to second order,
\be
\begin{split}
\vev{W_1} - \vev{W_1}_0 = & \sum_{l=1}^{\infty} \frac{(4\pi^2 b^2)^l}{(2l)!} \Biggl\{ - \sum_{n = 2}^\infty \frac{\zeta(2n-1)}{n}(-1)^n \sum_{k= 1}^{n-1} {2n \choose 2k} 2 {\cal V}(l,n-k) {\cal V}(k) \tilde{\lambda}_1^{l+n-k}\left( \tilde{\lambda}_1^k - \tilde{\lambda}_2^{k} \right) \\
& + \frac{1}{2} \sum_{n_1,n_2 = 2}^\infty  \frac{\zeta(2n_1-1)\zeta(2n_2-1)}{n_1 n_2}(-1)^{n_1+n_2} \sum_{k_i=1}^{n_i-1}{2n_1 \choose 2k_1} {2n_2 \choose 2k_2} \\ 
& \times \biggl[ 8 {\cal V}(l,n_1-k_1){\cal V}(k_1,n_2-k_2){\cal V}(k_2) \tlambda_1^{l+n_1-k_1}\left( \tilde{\lambda}_1^{k_1+n_2-k_2} + \tilde{\lambda}_2^{k_1+n_2-k_2} \right) \left( \tlambda_1^{k_2} - \tlambda_2^{k_2} \right)\\
& + 4 {\cal V}(l,n_1-k_1,n_2-k_2){\cal V}(k_1){\cal V}(k_2) \tilde{\lambda}_1^{l+n_1-k_1+n_2-k_2} \left( \tilde{\lambda}_1^{k_1} - \tilde{\lambda}_2^{k_1} \right) \left( \tilde{\lambda}_1^{k_2} - \tilde{\lambda}_2^{k_2} \right) \biggr] \Biggr\} \, , 
\end{split}
\ee
for which the corresponding rooted trees can be seen in figure (\ref{img:wl_trees}).

In Appendix B, we present the result of these sums up to order $\lambda^7$. We have checked that they reproduce the results of \cite{Mitev:2014yba, Mitev:2015oty}. Contrary to what happened for the free energy, the expectation value of this Wilson loop does not have nice properties under the exchange $\lambda_1 \leftrightarrow \lambda_2$. The reason is obvious, the Wilson loop is defined for one of the two gauge groups in the quiver, thus breaking the $\bZ_2$ symmetry. For this reason, let's consider the  linear combinations $\vev{W_1}\pm \vev{W_2}$, which were referred in \cite{Rey:2010ry}  as twisted and untwisted. These are symmetric and antisymmetric under the $\lambda_1 \leftrightarrow \lambda_2$ exchange, so we can introduce
\be
\vev{W_1}+\vev{W_2}-\vev{W_1}_0-\vev{W_2}_0 =(\lambda_1-\lambda_2)^2 \, w_+(\lambda_1,\lambda_2) \, ,
\ee

\be
\vev{W_1}-\vev{W_2}-\vev{W_1}_0+\vev{W_2}_0 = (\lambda_1-\lambda_2) \, w_-(\lambda_1,\lambda_2) \, ,
\ee
with $w_\pm$ symmetric under $\lambda_1 \leftrightarrow \lambda_2$. What is more, to the orders we have checked explicitly, again all the polynomials that appear in the expansion of $w_\pm$ have all their roots in the unit circle of the complex $\lambda_2/\lambda_1$ plane. We again conjecture that this is true for the polynomials generated by every tree.

For the polynomials that appear in $w_+(\lambda_1,\lambda_2)$, this would follow from our first conjecture if it is true. In particular, since in the previous section we proved the first conjecture for the simplest tree, it follows that it holds also for $w_+$, for the simplest tree. For $w_-$ the argument does not apply immediately, since $\vev{W_1}-\vev{W_2}-\vev{W_1}_0+\vev{W_2}_0 $ produces antipalindromic polynomials.

	\begin{figure}[h]
		\includegraphics[scale=1]{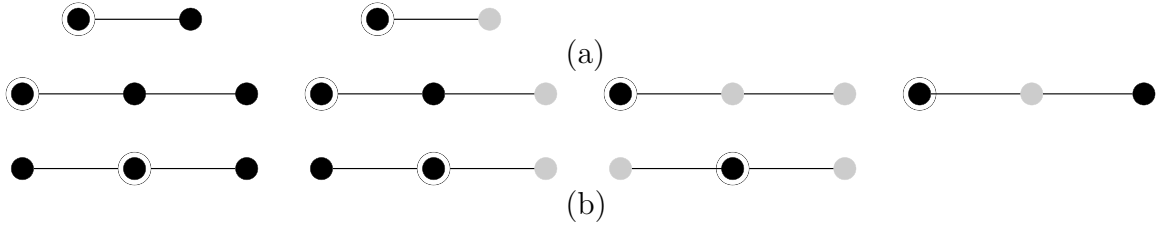}
		\put(-224,47){(a)}
		\put(-224,-10){(b)}
		\caption{Rooted trees corresponding to the Wilson loop in the large $N$, we see that inserting the operator selects from figure (\ref{img:treesb2}) trees with the same color as the operator that we are inserting, trees containing two different colors arise from interaction terms in (\ref{eq:eff_action}). $(a)$ Terms corresponding to ${\cal V}(l,n_1-k_1){\cal V}(k_1)$. $(b)$ Trees corresponding to ${\cal V}(l,n_1-k_1){\cal V}(k_1,n_2-k_2){\cal V}(k_2)$ and ${\cal V}(l,n_1-k_1,n_2-k_2){\cal V}(k_1){\cal V}(k_2)$.}
		\label{img:wl_trees}
	\end{figure}

To conclude, we can use these results to compute the one-point function of the energy-momentum tensor with these 1/2 BPS Wilson loops. This one-point function is fixed up to a coefficient $h_W$ \cite{Kapustin:2005py}, which can be obtained from the expectation value of the deformed Wilson loop $\vev{W_b}$ by the formula \cite{Fiol:2015spa, Bianchi:2019dlw}
\be
h_W= \frac{1}{12\pi^2}\, \partial_b \text{ ln } \vev{W_b} \vert_{b=1} \, .
\ee
finally, we can also compute the Bremsstrahlung function B \cite{Correa:2012at} using the relation $B=3h_W$ \cite{Fiol:2012sg, Lewkowycz:2013laa, Fiol:2015spa}, valid for any ${\cal N}=2$ superconformal field theory \cite{Bianchi:2018zpb}. The results we obtain agree with those of \cite{Mitev:2015oty}.

\acknowledgments
Research supported by Spanish MINECO under projects MDM-2014-0369 of ICCUB (Unidad de Excelencia ``Mar\'ia de Maeztu") and FPA2017-76005-C2-P, and by AGAUR, grant 2017-SGR 754.  J. M. M. is further supported by "la Caixa" Foundation (ID 100010434) with fellowship code LCF/BQ/IN17/11620067, and from the European Union's Horizon 2020 research and innovation programme under the Marie Sk{\l}odowska-Curie grant agreement No. 713673. A. R. F. is further supported by an FPI-MINECO fellowship. 

\appendix
\section{Planar free energy up to $6$th order}
Here we present the explicit form of the planar free energy in terms of $\lambda_i = \frac{\lambda_i}{16\pi^2}$
% \begin{equation}
% \begin{split}
% F_0(\lambda_1,\lambda_2) =& -3 \zeta_3 \left(\lambda_1 - \lambda_2\right)^2 -20 \zeta_5 \left(\lambda_1 - \lambda_2\right)^2\left(\lambda_1 + \lambda _2\right)   -70 \zeta_7 \left(\lambda_1 - \lambda_2\right)^2
%    \left(2 \lambda_1^2 + 3 \lambda_2 \lambda_1 + 2 \lambda_2^2\right) \\
%    & -84 \zeta_9 \left(\lambda_1 - \lambda_2\right)^2 \left(\lambda_1 + \lambda_2\right) \left(13 \lambda_1^2 + 10 \lambda_2 \lambda_1 + 13 \lambda_2^2\right) \\
%    &-154 \zeta_{11} \left(\lambda_1 - \lambda_2\right)^2 \left(61
%    \lambda_1^4 + 116 \lambda_2 \lambda_1^3 + 141 \lambda_2^2 \lambda_1^2 + 116
%    \lambda_2^3 \lambda_1 + 61 \lambda_2^4\right) \\
%    &+36 \zeta_3^2 \left(\lambda_1 - \lambda_2\right)^2 \left(\lambda _1^2+\lambda _2^2\right) + 240 \zeta_3 \zeta_5 \left(\lambda_1 - \lambda_2\right)^2 \left(\lambda_1 + \lambda_2\right)
%    \left(3 \lambda_1^2 -2 \lambda_2 \lambda_1 + 3 \lambda _2^2\right) \\
%    & + 840 \zeta_3 \zeta_7 \left(\lambda_1 - \lambda_2\right)^2 \left(8 \lambda_1^4 + 5 \lambda_2 \lambda_1^3 + 2 \lambda_2^2 \lambda_1^2 + 5 \lambda_2^3 \lambda_1 + 8 \lambda_2^4\right) \\
%    &+ 200 \zeta_5^2 \left(\lambda_1 - \lambda_2\right)^2
%    \left(19 \lambda _1^4+12 \lambda _2 \lambda
%    _1^3+\lambda _2^2 \lambda _1^2+12 \lambda _2^3
%    \lambda _1+19 \lambda _2^4\right) \\
%    &-144 \zeta _3^3 \left(\lambda _1-\lambda _2\right){}^2
%    \left(5 \lambda _1^4-2 \lambda _2 \lambda _1^3+6
%    \lambda _2^2 \lambda _1^2-2 \lambda _2^3 \lambda
%    _1+5 \lambda _2^4\right)+{\cal O}(\lambda^7)
% \end{split}
% \end{equation}

\begin{equation}
\begin{split}
{\cal F}_0(\lambda_1,\lambda_2) =  (\lambda_1 - \lambda_2)^2 \Bigl[ & - 3 \zeta_3 + 20 \zeta_5 \left(\lambda_1 + \lambda_2\right) - 70 \zeta_7 \left(2 \lambda_1^2 + 3 \lambda_1 \lambda_2 + 2 \lambda_2^2\right) \\
& + 84 \zeta_9  \left(\lambda_1 + \lambda_2\right) \left(13 \lambda_1^2 + 10 \lambda_1 \lambda_2 + 13 \lambda_2^2\right) \\
& - 154 \zeta_{11}  \left(61   \lambda_1^4 + 116 \lambda_1^3 \lambda_2 + 141 \lambda_1^2 \lambda_2^2 + 116 \lambda_1 \lambda_2^3 + 61 \lambda_2^4\right) \\
& + 36 \zeta_3^2 \left(\lambda_1^2 + \lambda_2^2\right) - 240 \zeta_3 \zeta_5 \left(\lambda_1 + \lambda_2\right) \left(3 \lambda_1^2 - 2 \lambda_1 \lambda_2 + 3 \lambda_2^2\right) \\
& + 840 \zeta_3 \zeta_7  \left(8 \lambda_1^4 + 5 \lambda_1^3 \lambda_2 + 2 \lambda_1^2 \lambda_2^2 + 5 \lambda_1 \lambda_2^3 + 8 \lambda_2^4\right) \\
& + 200 \zeta_5^2 \left(19 \lambda_1^4 + 12 \lambda_1^3 \lambda_2 + 4 \lambda_1^2 \lambda_2^2 + 12 \lambda_1 \lambda_2^3 + 19 \lambda_2^4\right) \\
& - 144 \zeta_3^3 \left(5 \lambda_1^4 - 2 \lambda_1^3 \lambda_2 + 6 \lambda_1^2 \lambda_2^2 - 2 \lambda_1 \lambda_2^3 + 5 \lambda_2^4\right) \Bigr]+{\cal O}(\lambda^7) .
\end{split}
\end{equation}

Up to the order we have explicitely checked, the polynomials have all unimodular roots.

\section{Wilson loop up to $\lambda^7$}

Here we present the explicit expansion of the circular Wilson loop corresponding to an insertion in the first node of the quiver; it is possible to obtain the insertion in the second node by making the change $\lambda_1 \leftrightarrow \lambda_2$. For simplicity, in the expansion we have set $b=1$ and $\lambda_i = \frac{\lambda_i}{16\pi^2}$. If one wishes to restore the powers of $b$ that appear in the perturbative expansion of $\vev{W_b}$ evaluated on $S^4$, one only needs to add in each term as many powers of $b$ as powers of $\pi$ there are.

\begin{equation}
\begin{split}
\vev{W_1} - \vev{W_1}_0 = \left(\lambda_1 - \lambda_2\right) \Bigl[ & - 24 \pi^2 \zeta_3 \lambda_1^2  - 32 \pi^4 \zeta_3 \lambda_1^3 - 16 \pi ^6 \zeta_3 \lambda_1^4 - \frac{64}{15} \pi^8 \zeta_3 \lambda_1^5 - \frac{32}{45} \pi^{10} \zeta_3 \lambda_1^6 \\
& + 80 \pi^2 \zeta_5 \lambda_1^2 \left(3 \lambda_1 + \lambda_2\right) + \frac{80}{3} \pi^4 \zeta_5 \lambda_1^3 \left(13 \lambda_1 + 4 \lambda_2\right) \\
& + \frac{32}{3} \pi^6 \zeta_5 \lambda_1^4 \left(17 \lambda_1 + 5 \lambda_2\right) + \frac{64}{9} \pi^8 \zeta_5 \lambda_1^5 \left(7 \lambda_1 + 2 \lambda_2\right) \\
& - 280 \pi^2 \zeta_7 \lambda_1^2 \left(8 \lambda_1^2 + 5 \lambda_1 \lambda_2 + \lambda_2^2\right) \\
& - \frac{112}{3} \pi^4 \zeta_7 \lambda_1^3 \left(91 \lambda_1^2 + 55 \lambda_1 \lambda_2 + 10 \lambda_2^2\right) \\
& - \frac{112}{3} \pi^6 \zeta_7 \lambda_1^4 \left(49 \lambda_1^2 + 29 \lambda_1 \lambda_2 + 5 \lambda_2^2\right) \\
& + 336 \pi^2 \zeta_9 \lambda_1^2 \left(5 \lambda_1 + \lambda_2\right) \left(13 \lambda_1^2 + 8 \lambda_1 \lambda_2 + 3 \lambda_2^2\right) \\
& + 672 \pi^4 \zeta_9 \lambda_1^3 \left(51 \lambda_1^3 + 41 \lambda_1^2 \lambda_2 + 17 \lambda_1 \lambda_2^2 + 2 \lambda_2^3\right) \\
& -3696 \pi^2 \zeta_{11} \lambda_1^2 \left(61 \lambda_1^4 + 56 \lambda_1^3 \lambda_2 + 36 \lambda_1^2 \lambda_2^2 + 11 \lambda_1 \lambda_2^3 + \lambda_2^4\right) \\
& + 288 \pi^2 \zeta_3^2 \lambda_1^2 \left(2 \lambda_1^2 - \lambda_1 \lambda_1 + \lambda_2^2\right) \\
& + 192 \pi^4 \zeta_3^2 \lambda_1^3 \left(5 \lambda_1^2 - 3 \lambda_1 \lambda_2 + 2 \lambda_2^2\right) \\
& + 192 \pi^6 \zeta_3^2 \lambda_1^4 \left(3 \lambda_1^2 - 2 \lambda_1 \lambda_2 + \lambda_2^2\right) \\
& - 960 \pi^2 \zeta_3 \zeta_5 \lambda_1^2 \left(15 \lambda_1^3 - 5 \lambda_1^2 \lambda_2 + \lambda_1 \lambda_2^2 + 5 \lambda_2^3\right) \\
& - 320 \pi^4 \zeta_3 \zeta_5 \lambda_1^3 \left(77 \lambda_1^3 - 32 \lambda_1^2 \lambda_2 + \lambda_1 \lambda_2^2 + 20 \lambda_2^3\right) \\
& + 3360 \pi^2 \zeta_3 \zeta_7 \lambda_1^2 \left(48 \lambda_1^4 - 7 \lambda_1^3 \lambda_2 - 7 \lambda_1^2 \lambda_2^2 + 11  \lambda_1 \lambda_2^3 + 11 \lambda_2^4\right) \\
& + 1600 \pi^2 \zeta_5^2 \lambda_1^2 \left(57 \lambda_1^4 - 8 \lambda_1^3 \lambda_2 - 10 \lambda_1^2 \lambda_2^2 + 14 \lambda_1 \lambda_2^3 + 13 \lambda_2^4\right) \\
& - 3456 \pi^2 \zeta_3^3 \lambda_1^2 \left(5 \lambda_1^4 - 5 \lambda_1^3 \lambda_2 + 5 \lambda_1^2 \lambda_2^2 - 3 \lambda_1 \lambda_2^3 + 2 \lambda_2^4\right) \Bigr] .
\end{split}
\label{res:wl1_o7}
\end{equation}
Note that we are inserting the operator in only one of the two nodes of the quiver thus breaking the $\bZ_2$ invariance of the theory. This is the reason why the vev (\ref{res:wl1_o7}) does not exhibit the same properties as the free energy. It is possible to retain the $\bZ_2$ invariance if we consider the sum and the difference, 
%The expectation value of the Wilson loop associated to a node does not transform nicely under the $\bZ_2$ symmetry of the quiver theory. For this reason, consider the sum and the difference,
%\be
%\vev{W_1}+\vev{W_2}-2\vev{W}^{{\cal N}=4} =(\lambda_1-\lambda_2)^2 w_+(\lambda_1,\lambda_2)
%\ee
%
%\be
%\vev{W_1}-\vev{W_2}= (\lambda_1-\lambda_2) w_-(\lambda_1,\lambda_2)
%\ee
for the case of the sum we have
\be
\begin{split}
%\vev{W_1} + \vev{W_2} = \left(\lambda_1 - \lambda_2\right)^2 
w_+(\lambda_1,\lambda_2)=
\Bigl[ & - 24 \pi^2 \zeta_3  \left(\lambda_1 + \lambda _2\right) - 32 \pi^4 \zeta_3 \left(\lambda_1^2 + \lambda_1 \lambda_2 + \lambda_2^2\right) \\
& - 16 \pi ^6 \zeta_3 \left(\lambda_1 + \lambda_2\right) \left(\lambda_1^2 + \lambda_2^2\right) - \frac{64}{15} \pi^8 \zeta_3 \left(\lambda_1^4 + \lambda_1^3 \lambda_2 + \lambda_1^2 \lambda_2^2 + \lambda_1 \lambda_2^3 + \lambda_2^4\right) \\
& - \frac{32}{45} \pi^{10} \zeta_3 \left(\lambda_1 + \lambda_2\right) \left(\lambda_1^4 + \lambda_1^2 \lambda_2^2 + \lambda_2^4\right) + 80 \pi^2 \zeta_5 \left(3 \lambda_1^2 + 4 \lambda_1 \lambda_2 + 3 \lambda _2^2\right) \\
& + \frac{80}{3} \pi^4 \zeta_5 \left(\lambda_1 + \lambda_2\right) \left(13 \lambda_1^2 + 4 \lambda_1 \lambda_2 + 13 \lambda_2^2\right) \\
& + \frac{32}{3} \pi^6 \zeta_5 \left(17 \lambda_1^4 + 22 \lambda_1^3 \lambda_2 + 22 \lambda_1^2 \lambda_2^2 + 22 \lambda_1 \lambda_2^3 + 17 \lambda_2^4\right) \\
& + \frac{64}{9} \pi^8 \zeta_5 \left(\lambda_1 + \lambda_2\right) \left(7 \lambda_1^4 + 2 \lambda_1^3 \lambda_2  + 7 \lambda_1^2 \lambda_2^2 + 2 \lambda_1 \lambda_2^3 + 7 \lambda_2^4\right) \\
& - 280 \pi^2 \zeta_7 \left(\lambda_1 + \lambda_2\right) \left(8 \lambda_1^2 +5 \lambda_1 \lambda_2 + 8 \lambda_2^2\right) \\
& - \frac{112}{3} \pi^4 \zeta_7 \left(91 \lambda_1^4 + 146 \lambda_1^3 \lambda_2 + 156 \lambda_1^2 \lambda_2^2 + 146 \lambda_1 \lambda_2^3 + 91 \lambda_2^4\right) \\
& - \frac{112}{3} \pi^6 \zeta_7 \left(\lambda_1 + \lambda_2\right) \left(49 \lambda_1^4 + 29 \lambda_1^3 \lambda_2 + 54 \lambda_1^2 \lambda_2^2 + 29 \lambda_1 \lambda_2^3 + 49 \lambda_2^4\right) \\
& + 336 \pi^2 \zeta_9 \left(65 \lambda_1^4 + 118 \lambda_1^3 \lambda_2 + 138 \lambda_1^2 \lambda_2^2 + 118 \lambda_1 \lambda_2^3 + 65 \lambda_2^4\right)\\
& + 672 \pi^4 \zeta_9 \left(\lambda_1 + \lambda_2\right) \left(51 \lambda_1^4 + 41 \lambda_1^3 \lambda_2 + 68 \lambda_1^2 \lambda_2^2 + 41 \lambda_1 \lambda_2^3 + 51 \lambda_2^4\right)  \\
& - 3696 \pi^2 \zeta_{11} \left(\lambda_1 + \lambda_2\right) \left(61 \lambda_1^4 + 56 \lambda_1^3 \lambda_2 + 96 \lambda_1^2 \lambda_2^2 + 56 \lambda_1 \lambda_2^3 + 61 \lambda_2^4\right) \\
& + 288 \pi^2 \zeta_3^2 \left(\lambda_1 + \lambda_2\right) \left(2 \lambda_1^2 - \lambda_1 \lambda_2 + 2 \lambda_2^2\right) \\
& + 192 \pi^4 \zeta_3^2 \left(5 \lambda_1^4 + 2 \lambda_1^3 \lambda_2 + 4 \lambda_1^2 \lambda_2^2 + 2 \lambda_1 \lambda_2^3 + 5 \lambda_2^4\right) \\
& + 192 \pi^6 \zeta_3^2 \left(\lambda_1 + \lambda_2\right) \left(3 \lambda_1^4 - 2 \lambda_1^3 \lambda_2 + 4 \lambda_1^2 \lambda_2^2 - 2 \lambda_1 \lambda_2^3 + 3 \lambda_2^4\right) \\
& - 960 \pi^2 \zeta_3 \zeta_5 \left(15 \lambda_1^4 + 10 \lambda_1^3 \lambda_2 + 6 \lambda_1^2 \lambda_2^2 + 10 \lambda_1 \lambda_2^3 + 15 \lambda_2^4\right) \\
& - 320 \pi^4 \zeta_3 \zeta_5 \left(\lambda_1 + \lambda_2\right) \left(77 \lambda_1^4 - 32 \lambda_1^3 \lambda_2 + 78 \lambda_1^2 \lambda_2^2 - 32 \lambda_1 \lambda_2^3 + 77 \lambda_2^4\right) \\
& + 3360 \pi^2 \zeta_3 \zeta_7 \left(\lambda_1 + \lambda_2\right) \left(48 \lambda_1^4 - 7 \lambda_1^3 \lambda_2 + 30 \lambda_1^2 \lambda_2^2 - 7 \lambda_1 \lambda_2^3 + 48 \lambda_2^4\right) \\
& + 1600 \pi^2 \zeta_5^2 \left(\lambda_1 + \lambda_2\right) \left(57 \lambda_1^4 - 8 \lambda_1^3 \lambda_2 + 34 \lambda_1^2 \lambda_2^2 - 8 \lambda_1 \lambda_2^3 + 57 \lambda_2^4\right) \\
& - 3456 \pi^2 \zeta_3^3 \left(\lambda_1 + \lambda_2\right) \left(5 \lambda_1^4 - 5 \lambda_1^3 \lambda_2 + 8 \lambda_1^2 \lambda_2^2 - 5 \lambda_1 \lambda_2^3 + 5 \lambda_2^4\right) \Bigr] .
\end{split}
\ee

For the case of the difference we have
\be 
\begin{split}
%\vev{W_1} - \vev{W_2} = \left(\lambda_1 - \lambda_2\right) 
w_-(\lambda_1,\lambda_2)=\Bigl[ & - 24 \pi^2 \zeta_3 \left(\lambda_1^2 + \lambda_2^2\right) - 32 \pi^4 \zeta_3 \left(\lambda_1^3 + \lambda_2^3\right) - 16 \pi^6 \zeta_3 \left(\lambda_1^4 + \lambda_2^4\right) \\
& - \frac{64}{15} \pi ^8 \zeta_3 \left(\lambda_1^5 + \lambda_2^5\right) - \frac{32}{45} \pi^{10} \zeta_3 \left(\lambda_1^6 + \lambda_2^6\right) \\
& + 80 \pi^2 \zeta_5 \left(\lambda_1 + \lambda_2\right) \left(3 \lambda_1^2 - 2 \lambda_1 \lambda_2 + 3 \lambda_2^2\right) \\
& + \frac{80}{3} \pi^4 \zeta_5 \left(13 \lambda_1^4 + 4 \lambda_1^3 \lambda_2 + 4 \lambda_1 \lambda_2^3 + 13 \lambda_2^4\right) \\
& + \frac{32}{3} \pi^6 \zeta_5 \left(\lambda_1 + \lambda_2\right) \left(17 \lambda_1^4 - 12 \lambda_1^3 \lambda_2 + 12 \lambda_1^2 \lambda_2^2 - 12 \lambda_1 \lambda_2^3 + 17 \lambda_2^4\right) \\
& + \frac{64}{9} \pi^8 \zeta_5 \left(7 \lambda_1^6 + 2 \lambda_1^5 \lambda_2 + 2  \lambda_1\lambda_2^5 + 7 \lambda_2^6\right) \\
& - 280 \pi^2 \zeta_7 \left(8 \lambda_1^4 + 5 \lambda_1^3 \lambda_2 + 2 \lambda_1^2 \lambda_2^2 + 5 \lambda_1 \lambda_2^3 + 8 \lambda_2^4\right) \\
& - \frac{112}{3} \pi^4 \left(\lambda_1 + \lambda_2\right) \left(91 \lambda_1^4 - 36 \lambda_1^3 \lambda_2 + 46 \lambda_1^2 \lambda_2^2 - 36 \lambda_1 \lambda_2^3 + 91 \lambda_2^4\right) \\
& - \frac{112}{3} \pi^6 \zeta_7 \left(49 \lambda_1^6 + 29 \lambda_1^5 \lambda_2 + 5 \lambda_1^4 \lambda_2^2 + 5 \lambda_1^2 \lambda_2^4 + 29 \lambda_1 \lambda_2^5 + 49 \lambda_2^6\right) \\
& + 336 \pi^2 \zeta_9 \left(\lambda_1 + \lambda_2\right) \left(65 \lambda_1^4 - 12 \lambda_1^3 \lambda_2 + 38 \lambda_1^2 \lambda_2^2 - 12 \lambda_1 \lambda_2^3 + 65 \lambda_2^4\right) \\
& + 672 \pi^4 \zeta_9 \left(51 \lambda_1^6 + 41 \lambda_1^5 \lambda_2 + 17 \lambda_1^4 \lambda_2^2 + 4 \lambda_1^3 \lambda_2^3 + 17 \lambda_1^2 \lambda_2^4 + 41 \lambda_1 \lambda_2^5 + 51 \lambda_2^6\right) \\
& - 3696 \pi^2 \zeta_{11} \left(61 \lambda_1^6 + 56 \lambda_1^5 \lambda_2 + 37 \lambda_1^4 \lambda_2^2 + 22 \lambda_1^3 \lambda_2^3 + 37 \lambda_1^2 \lambda_2^4 + 56 \lambda_1 \lambda_2^5 + 61 \lambda_2^6\right) \\
& + 288 \pi^2 \zeta_3^2 \left(2 \lambda_1^4 - \lambda_1^3 \lambda_2 + 2 \lambda_1^2 \lambda_2^2 - \lambda_1 \lambda_2^3 + 2 \lambda_2^4\right) \\
& + 192 \pi^4 \zeta_3^2 \left(\lambda_1 + \lambda_2\right) \left(\lambda_1^2 + \lambda_2^2\right) \left(5 \lambda_1^2 - 8 \lambda_1 \lambda_2 + 5 \lambda_2^2\right) \\
& + 192 \pi^6 \zeta_3^2 \left(3 \lambda_1^6 - 2 \lambda_1^5 \lambda_2 + \lambda_1^4 \lambda_2^2 + \lambda_1^2 \lambda_2^4 - 2 \lambda_1 \lambda_2^5 + 3 \lambda_2^6\right) \\
& - 960 \pi^2 \zeta_3 \zeta_5 \left(\lambda_1 + \lambda_2\right) \left(15 \lambda_1^4 - 20 \lambda_1^3 \lambda_2 + 26 \lambda_1^2 \lambda_2^2 - 20 \lambda_1 \lambda_2^3 + 15 \lambda_2^4\right) \\
& - 320 \pi^4 \zeta_3 \zeta_5 \left(77 \lambda_1^6 - 32 \lambda_1^5 \lambda_2 + \lambda_1^4 \lambda_2^2 + 40 \lambda_1^3 \lambda_2^3 + \lambda_1^2 \lambda_2^4 - 32 \lambda_1 \lambda_2^5 + 77 \lambda_2^6\right) \\ 
& + 3360 \pi^2 \zeta_3 \zeta_7 \left(48 \lambda_1^6 - 7 \lambda_1^5 \lambda_2 + 4  \lambda_1^4\lambda_2^2 + 22 \lambda_1^3 \lambda_2^3 + 4 \lambda_1^2 \lambda_2^4 - 7 \lambda_1 \lambda_2^5 + 48 \lambda_2^6\right) \\
& + 1600 \pi^2 \zeta_5^2 \left(57 \lambda_1^6 - 8 \lambda_1^5 \lambda_2 + 3 \lambda_1^4 \lambda_2^2 + 28 \lambda_1^3 \lambda_2^3 + 3 \lambda_1^2 \lambda_2^4 - 8 \lambda_1 \lambda_2^5 + 57 \lambda_2^6\right) \\
& - 3456 \pi^2 \zeta_3^3 \left(5 \lambda_1^6 -5 \lambda_1^5 \lambda_2 + 7 \lambda_1^4 \lambda_2^2 - 6 \lambda_1^3 \lambda_2^3 + 7 \lambda_1^2 \lambda_2^4 - 5 \lambda_1 \lambda_2^5 + 5 \lambda_2^6\right) \Bigr] .
\end{split}
\ee

The series $w_\pm (\lambda_1,\lambda_2)$ are symmetric. At the considered orders, the polynomials that appear also have all unimodular roots.

\bibliographystyle{JHEP}

\end{document}